\documentclass[12pt]{article}
\usepackage{a4,latexsym,graphicx,color}
\usepackage[numbers]{natbib}
\usepackage{amsmath}
\usepackage{amssymb}
\usepackage{amsfonts}
\usepackage{epsfig}
\usepackage{url}

\newcommand{\modif}[1]{{\color{black} #1}} 

\begin{document}

\title{Unified presentation of  four fundamental inequalities}

\author{Joseph Lajzerowicz$^1$, Roland Lehoucq$^2$, Fran\c{c}ois Graner$^3$}

\date{\today}

\maketitle

{\small $^{1}$Retired from Laboratoire Interdisciplinaire de Physique, CNRS UMR 5588, Universit\'e Grenoble Alpes, Saint-Martin d'H\`eres, France. \par $^{2}$Laboratoire AIM-Paris-Saclay, CEA/DSM/Irfu, CNRS UMR 7158, Universit\'e Paris Diderot, CEA Saclay, Gif-sur-Yvette, France. \par $^{3}$Laboratoire Mati\`ere et Syst\`emes Complexes, CNRS UMR 7057, Universit\'e Paris Diderot, Paris, France.}

\begin{abstract}
We suggest an unified presentation to teach fundamental constants to graduate students, by introducing four lower limits to observed phenomena.  
The reduced Planck constant $\hbar$ is the lowest \modif{classically definable} action. The inverse of invariant speed, $s$, is the lowest observable slowness. The Planck time, $t_{\rm P}$, is the lowest observable time scale. The Boltzmann constant,  $k$, determines the lowest \modif{coherent} degree of freedom\modif{; w}e recall a Einstein criterion on the \modif{fluctuations} of small thermal systems \modif{and show that it} has far-reaching implications, such as demonstrating the relations between critical exponents. 
Each of these four fundamental limits enters in an inequality, which marks a horizon of the universe we can perceive. This compact presentation can resolve some difficulties encountered when trying to defining the epistemologic status of these constants, and emphasizes their useful role in shaping our intuitive vision of the universe. 
\end{abstract}

\section{Fundamental constants}

The so-called ``natural" \cite{Okun2002,Barrow2003}, ``universal" \cite{Cohen1991,nist} or ``fundamental" 
\cite{Uzan2003,Uzan2005,Uzan2008,BIPM} constants have a controversed status, and even their number is controversial. We will not attempt at reviewing the huge literature on the subject (see e.g. \cite{Uzan2003} and references therein). It suffices here to mention that in books, articles, seminars or classes, one encounters for instance the four following constants. 

\modif{
The reduced Planck constant, $\hbar$, relates the frequency and energy exchanged by an oscillator. 
The invariant speed, $c$, or speed of light in vacuum, relates invariant mass and rest energy. 
The constant of gravitation, $G$, relates the masses of interacting objects with their Newtonian interaction force, or  the relativistic energy-momentum tensor with the space-time curvature. 
The Boltzmann constant, $k$, relates the temperature of a system and the energy of its thermal fluctuations, or  the thermal fluctuations of a system and its dissipation, or the entropy of a system and the number of  its accessible microscopic states.
}

These constants have a dimension. They can be all taken as  equal to one by redefining the system of units. Thus the importance of these constants does not lie in their actual value\modif{, but rather in the fact that they enter in important equations which relate very different domains \cite{JMLL1977}}. We do not discuss here their possible variation, especially given the fact that $c$ is fixed and, with the second, defines the meter in the SI units system. In 2018 $\hbar$ and $k$ will be fixed too \cite{BIPM,Newell2014,Knotts2016}  in order to define the kilogram and the \modif{k}elvin. More generally,  the number of the constants, their combinations and the prefactors are conventional choices. 

Let us immediately make clear that ``coupling constants" are different. The  coupling constants are dimensionless, and far more numerous. They include for instance Sommerfeld's  fine-structure constant, the strong coupling constant, and other coupling constants which describe interactions. The fact that their value is smaller or larger than 1 has an intrinsic meaning and defines different regimes. Thus coupling constants and fundamental constants should not be confused.

\section{Fundamental limits and inequalities}

\subsection{A set of four limits}

We \modif{suggest} that the interest of the  fundamental constants
lies in their role  in an \modif{\it inequality}, which \modif{evidences} essential  limits of the universe we can perceive.
\modif{Some of these limits have been made intuitive by Gamow's tales \cite{Gamow1946}, where a character called ``Mr Tompkins" undergoes surprising adventures because the values of the fundamental constants change. In fact, these constants} delimitate the realm of scientific studies, whether current or future, from that of metaphysical speculations.  
Far from these limits is the realm of classical physics. 
Thus, it might be appropriate to define them as ``fundamental limits" entering ``fundamental inequalities". This definition might turn more robust to technical progresses than definitions \cite{Cohen1991} based on  horizons related with our current capacity to perceive the universe. 

Since the choice of  fundamental limits is arbitrary, we are entitled to motivate our choice by pedagogy and elegance, or even by reference to poetry ({\it les quatre horizons qui crucifient le monde}, the four horizons that crucify the world \cite{Jammes1913}).
 We select four limits which \modif{each} enter in a fundamental inequality (see Table \ref{tab:four_limits}): 

\modif{
\begin{itemize}
\item 
The Planck constant  over $2\pi$, $\hbar$, is the lowest value of classically definable action, that is, angular momentum, or product of energy with time. 
\item  The invariant speed, $c$, is the maximal physical speed, reached by massless objects and only by them, and the maximal speed for causal signals.
In order to introduce a lower rather than an upper limit, we will use its inverse,  $s=c^{-1}$, which can be called the ``invariant slowness", where the useful notion of slowness is defined and discussed below.
\item The Planck frontier  is the lowest observable scale, where ``scale" can refer to time, length or energy scale, as discussed below.
\item  The Boltzmann constant $k$ enters in the definition of the lowest coherent degree of freedom.
\end{itemize}
}

The Planck constant is probably the most consensual one 
\modif{(the division of $h$ by  $2\pi$, introduced by Dirac, is more than a handy convention \cite{JMLL2002})}.
We now comment briefly the notion of slowness (section \ref{sec:slowness}) and the choice of the Planck \modif{frontier}  (section \ref{sec:planck})\modif{, then in more details the physical role of Boltzmann constant and its applications} (section \ref{sec:Boltz}).

\begin{table}[h]
\begin{center}
\begin{tabular}{|c|c|c|c|c|}
\hline
{\it constant}Ê & $\hbar$  & $s$ & $t_{\rm P}$ & $k$  \\
\hline
{\it value}Ê &   1.054 571 800(13) & 3.335 640 952 \ldots   &  5.391 16(13)   &  1.380 648 52(79)  \\
{\it \& unit}Ê &    $\times$ 10$^{-34}$ J s & $\times$ 10$^{-9}$ s m$^{-1}$  & $\times$ 10$^{-44}$ s  &    $\times$ 10$^{-23}$ J K$^{-1}$ \\
\hline
{\it limit}Ê & \modif{action} & slowness & scale & \modif{coherence} \\
\hline
{\it limit}Ê & Heisenberg & Einstein & Planck & Einstein \\
{\it name}Ê & principle & causality & \modif{frontier}  & \modif{fluctuations} \\
\hline
{\it theoretical}Ê & quantum & special  & scale & statistical \\
{\it domain}Ê& physics & relativity & relativity & physics \\
\hline
\end{tabular}
\caption{Four fundamental limits and associated constants. These values and their uncertainty (in brackets) have been listed in 2014 \cite{nist}. The value of $s$ is fixed and thus without uncertainty. With the new base for SI units in 2018, values for $\hbar$ and $k$ become fixed too \cite{BIPM,Newell2014,Knotts2016}. }
\label{tab:four_limits}
\end{center}
\vspace{-0.6cm}
\end{table}

\subsection{Slowness}
 \label{sec:slowness}

Slowness is the inverse of speed. This interesting notion  would deserve a separate article. 
Teaching essential limits and teaching slowness mutually reinforce each other and could be done in any order. For instance, one could use Table  \ref{tab:four_limits} as a mean to introduce the notion and usefulness of slowness. Or, after having explained slowness in daily life, it could be used to introduce naturally Table  \ref{tab:four_limits}.

\modif{In daily life, few persons care for the addition of speeds; when you walk inside a plane, boat or train, or on a treadmill, you seldom need to estimate your velocity with respect to the ground (you can leave that as an exercise for physics students). On the opposite, what we routinely use in daily life is the addition of durations required to go from a given place to another one, and thus  of the distances multiplied by slowness.} 
Sportspersons like runners \modif{claim their times rather than their speeds; they} record ``minutes per kilometer": 
this quantity is additive\modif{; i}ts time integral has a  physical meaning, which is the total duration.
Sismologists use slowness because they compose different materials with different densities and hence different acoustic indices  \cite{Kennett1981}. 

Slownesses compose according to the relativistic law of composition, just as the speeds do. 
Slowness might be more fundamental than speed: it is analogous with optical indices and  is part of Fermat, Huygens and Maupertuis least action principles \cite{Kimball1998}. For instance for  ray refraction at an interface, the Huygens construction for rays and wavefronts duality uses slowness and transposes in the Fourier space (``dispersion surface", i.e. surface of wavevectors) what the Descartes construction  performs with speed in real space.

\modif{Following Malus, given an initial position $\vec{r}_0$, one can define  the ``duration" function $d\left(\vec{r}\right) = T\left(\vec{r}_0,\vec{r}\right)$  which is the time $T$ required for an object starting from $\vec{r}_0$ to reach the position $\vec{r}$. The gradient of the function $d$ is a vector, with the unit of an inverse speed, which defines the vectorial slowness: $\vec{\cal L} = \nabla d$. For instance, when a point source emits light, the wave surfaces are sets of points with same $d$; then $\vec{\cal L}$ is perpendicular to these wave surfaces and parallel to the wave vector.}
The scalar product of slowness and speed vectors is \modif{$\vec{\cal L}.\vec{v}=1$  \cite{Kimball1998}.} 
Optimising a boat trajectory requires to direct the slowness \modif{vector $\vec{\cal L}$} (rather than the speed $\vec{v}$) towards the goal \cite{Kimball1998}.

Here \modif{$s = c^{-1}$ is the slowness of light in vacuum, the minimal physical speed, reached by massless objects and only by them, and the minimal speed for causal signals.
Without any reference to these classical significations, $s$ can also be defined}
as the invariant speed in changes of inertial frames \cite{LevyLeblond1976,Pelissetto2015}.
Note that
$s^2$ is the single parameter left free by the symmetries in the derivation of L\'evy-Leblond \cite{LevyLeblond1976}.

\subsection{Planck \modif{frontier} }
\label{sec:planck}

The Planck \modif{frontier}  can be associated with a time scale $t_{\rm P} = \left(\hbar G / c^5\right)^{1/2}$, a length scale $\ell_{\rm P}= \left(\hbar G / c^3\right)^{1/2}$, or an energy scale $E_{\rm P}= \left(\hbar c^5/G\right)^{1/2}$.
These Planck time, length and energy arise immediately from dimensional analysis from the Planck constant, even within Newtonian gravitation. However,  their modern interpretation as a \modif{frontier resulting from} an inequality is due to Bronstein, in 1936, and is rooted in general relativity \cite{Bronstein1936a,Bronstein1936b,Gorelik1994,Rovelli2015}, as follows.

The Planck \modif{frontier}  characterises the competition between gravitational  binding  and limitations to confinment in quantum mechanics. Beyond the Planck \modif{frontier}, quantum fluctuations of gravitational field are so large that space-time is no longer a continuous differential variety. As Rovelli and Vidotto write it \cite{Rovelli2015}, Bronstein's above original argument is that space and time are necessarily ill-defined in quantum gravity. The solution is to accept that observables  do not resolve space and time more finely than Planck scale. This  forces the connection between gravity and geometry and is the core of modern attempts towards quantum gravity.

Equivalently, an object which Compton length is equal to $\ell_{\rm P}$ is a mini blackhole.
Quantitatively, this amounts to writing that for an object of mass $m$, the Schwarzschild radius is $r=2Gm/c^2$, and the object is at most confined over a size equal to its Compton length, $r= \hbar/2mc$.
Eliminating $m$ yields $r^2=\hbar G/c^3$, and hence the Planck size, $r=\ell_{\rm P}$; the Planck time $t_{\rm P}$ is derived immediately. Alternatively, eliminating $r$ yields the Planck mass; the Planck energy $E_{\rm P}$ is derived immediately.
This derivation uses concepts that predate Bronstein (Schwarzschild radius in 1916, Compton effect in 1923). It can even be formulated in a Newtonian framework, where the escape speed  is $c$ if the object has a size $r=2Gm/c^2$: surprisingly,  even the prefactor is correct.

There could be arguments to favor  the Planck length $\ell_{\rm P}$\modif{;
f}or instance, scale relativity theories invoke  an invariant scale, which is more often presented as a length than as a duration \cite{Nottale1993,Rovelli2015}.
But we suggest to use the Planck time, $t_{\rm P}$, which  is the lowest observable time scale.
In fact, we find this choice is pedagogical for graduate students, because the time is a true scalar while space is more associated with vectors; and 
 because Planck energy is a upper limit rather than a lower one. 
Also, note that in popular science  the Big Bang  scenario is usually more described in terms of time  than of length or of energy. Finally,  since 1983 the  durations are conventionnal\modif{l}y defined as more fundamental quantities than the  lengths (which are now derived from durations through $c$).

\section{Boltzmann constant}
\label{sec:Boltz}

It has been argued that $k$ is not fundamental in the sense that its value can change without  fundamental affecting physical phenomena \cite{Uzan2005}. 
In May 2017, the NIST website lists the Boltzmann constant simply as a ``frequently used" constant. What it lists as ``fundamental" constant is rather the Planck temperature, $T_{\rm P}$, defined as the ratio of the Planck energy to the Boltzmann constant,  $E_{\rm P}/k$ \cite{nist}. 

On the opposite, there exist arguments in favour of its fundamental status. If a modern Gamow wrote the story of a  country where $k$ is high, animals would become Brownian, especially smaller ones, and hunting  might turn as difficult as in a quantum country  \cite{Gamow1946}.
With the new base for SI units in 2018,  $k$ will become fixed  \cite{BIPM,Newell2014,Knotts2016}. It enters into a fundamental inequality, as we will now explain in detail.

\subsection{Gibbs and Einstein estimation of fluctuations}

The energy $E$ of a system at temperature $T$ fluctuates. Gibbs  \cite{Gibbs1902} showed that the variance $\sigma^2= \left< E^2\right> - \left< E\right>^2$ of this energy obeys:
\begin{equation}
\sigma^2  = kT^2 \frac{\partial E}{\partial T}
\label{Gibbs}
\end{equation}
where $\left< E\right>$ is the average over fluctuations, and $\partial E/\partial T$ is the system's heat capacity corresponding to the existing constraints (such as constant pressure or constant volume). 
We check that both sides of eq. (\ref{Gibbs}) are insensitive to an additive constant in the energy. 

Eq. (\ref{Gibbs}) is largely independent from the hypotheses used to derive it. 
For instance, Einstein later re-demonstrated it independently  \cite{Einstein1904,Peliti2017}. 
His demonstration  goes as follows (he denoted by $2\kappa$ what we note $k$, and we refer to a version where typographic mistakes have been corrected \cite{Einstein1989}). Since $ \left< E\right> = \int E \; {\cal P}(E) dE$ where the probability of the value $E$ is $ {\cal P}(E) \propto \exp\left(-E/kT \right) $, one can simply write:
\begin{equation}
 \int \left(E - \left< E\right> \right)  \exp^{-\frac{E}{kT}} dE = 0
\label{Einstein}
\end{equation}
Differentiating the \modif{left hand side} of eq. (\ref{Einstein}) with respect to $T$ yields eq. (\ref{Gibbs}).

\subsection{A fundamental inequality: Einstein criterion}

Gibbs remarks that the energy is extensive, so that $E$ is proportional to the volume $V$ of the system under consideration; thus $\sigma$ goes as $V^{1/2}$ and is much smaller that $E$ in the limit of large volumes. 
Einstein similarly remarks that, on the opposite limit, if the system size $V$ is small enough,  $\sigma$ becomes comparable to $E$, or even larger than $E$ \modif{\cite{Einstein1904}. Without expliciting his thought, he concludes that $k$ plays a role in the system stability. Einstein criterion states there is a critical volume such that:} 
\begin{equation}
E > \sigma
\label{einstein_crit}
\end{equation}
It can be written, using eq. (\ref{Gibbs}), as a fundamental inequality involving $k$: 
\begin{equation}
\left(\frac{E}{T}\right)^2 \left(\frac{\partial E}{\partial T}\right)^{-1} > k
\label{fundamental}
\end{equation}
Using the energy density  $e= E/V$, eq. (\ref{fundamental}) can be equivalently rewritten as:
\begin{equation}
V > V_s
\label{volume_ineq}
\end{equation}
where
\begin{equation}
V_s = k \frac{\partial e}{\partial T} \left(\frac{e}{T}\right)^{-2} 
\label{eq:crit_vol}
\end{equation}
and $\partial e/\partial T$ is the volume-specific heat capacity.
The physical interpretation of $V_s$ is the volume which energy is equal to its \modif {thermal fluctuations (which might impact on its stability)}; or a coherent volume, which defines the effective thermodynamical degree of freedom.

\subsection{Einstein's applications of his criterion}

Einstein \modif{remarks, without providing \modif{explanations}, that this criterion is difficult to apply in practice \cite{Einstein1904}; and he suggests one application, namely} to the blackbody.
He finds that $V_s^{1/3} T = 0.42 \; 10^{-3}$~m. Thus $V_s^{1/3}$ correctly yields the wavelength of the  blackbody radiation maximum (Wien's law), inversely proportional to the temperature, $\lambda_m = \sigma_W / T$, with even the good order of magnitude for the Wien constant, $\sigma_W = 0.2898 \; 10^{-3}$~m K$^{-1}$ (in his article, Einstein used the value 0.293). This mere 43\% overestimating is a good agreement and, given the generality of the hypotheses involved, it is probably not a coincidence, \modif{says} Einstein.

Alternatively, we note that $V_s^{1/3}$ and $\lambda_m$ have the same order of magnitude as the correlation length of fluctuations in a gas of photons at thermal equilibrium, $\hbar c/kT$ \cite{landau_statphys}.

As mentioned above, both sides of eq. (\ref{Gibbs}) are insensitive to an additive constant in the energy. However, eq. (\ref{einstein_crit}) does depend on an additive constant in the energy, and implicitly assumes that there is a reference state with zero energy. \modif{We suggest that one important difficulty in applying this criterion resides in the necessity to define without ambiguity such a reference energy. That might explain}  why Einstein applies it to the blackbody radiation: in this case, the reference energy is intrinsically and trivially zero, representing the space devoid of photons. 

\subsection{Other applications of Einstein's criterion}

\modif{We argue that o}ther applications of eq. (\ref{Gibbs}) include the determination of scaling laws in phase transitions.  \modif{B}oth sides of eq. (\ref{einstein_crit}) diverge, but they diverge together, and it is this concomitant divergence that we examine. 
\modif{This has been briefly done by one of us  (pp. 1211, 1214, 1215 of ref. \cite{Bastie1978}). We can rephrase the argument in a more general form, as follows.}
The critical exponent for energy density is $\alpha$ , defined by :
\begin{eqnarray}
e & \propto& t^{1-\alpha} \nonumber \\
\frac{\partial e}{\partial T} & \propto& t^{-\alpha}
\end{eqnarray}
where $t = (T-T_c)/T_c$ is the reduced temperature near the transition temperature $T_c$.
The correlation length $ \ell_c$ scales with an exponent $\nu$:
\begin{equation}
 \ell_c \propto t^{-\nu}
\end{equation}
So the correlation volume goes as $ \ell_c^D$, where $D$ is the dimension of space:
\begin{equation}
V \propto t^{-D\nu}
\end{equation}
Eqs. (\ref{Gibbs},\ref{eq:crit_vol}) yield $-D \nu  -\alpha = -2 D \nu + 2 (1-\alpha)$, i.e. : 
\begin{equation}
\alpha + D \nu = 2
\end{equation}
This is one of the scaling relation between critical exponents  \cite{landau_statphys}. They all can be derived similarly, in this simple, general, unified way which is  pedagogical for graduate students.

Replacing the energy with the magnetization, and the specific heat with the magnetic susceptibility, eq. (\ref{Gibbs}) yields the Ginzburg-Levanyuk inequality which marks the onset of the critical regime \modif{\cite{landau_statphys,Bastie1978}}. Similarly, the Debye length in ionic solutions can be interpreted as a length marking the competition  between potential energy (which groups the charges together) and fluctuations (entropic dispersion of the charges)\modif{; in this example, where the energy vanishes when $T$ goes to infinity, applying Einstein criterion is subtle}.

\section{Conclusion}

The name ``Bronstein cube" is sometimes given to the ``cube of physical theories", graphically representing $c$, $G$ and $\hbar$ as three axes  \cite{Gamow1928,Uzan2005}. This ``cube" becomes an hypercube if one also includes $k$ (see \cite{Okun2002} and references therein). We suggest that $\left(\hbar,s, t_{\rm P},k\right)$ could be the axes of a hypercube, which would encompass  the current set of physical theories \modif{and their fundamental limits}.

\section*{Acknowledgments}

We thank Marcel Vallade \modif{for the idea that fundamental constants correspond to inequalities,} Jean-Philippe Uzan for stimulating discussions, Boris Bolliet for references\modif{, Jean-Marc L\'evy-Leblond for critical reading of the manuscript}.

\end{document}